# 3D Virtual Reality vs. 2D Desktop Registration User Interface Comparison

Andreas Bueckle[1,*], Kilian Buehling[2], Patrick C. Shih[3], Katy Börner[1,4]

[1] Department of Intelligent Systems Engineering, Luddy School of Informatics, Computing, and Engineering, Indiana University, Bloomington, Indiana, United States of America

[2] Research Group Knowledge and Technology Transfer, Fakultät Wirtschaftswissenschaften, Technische Universität Dresden, Germany

[3] Department of Informatics, Luddy School of Informatics, Computing, and Engineering, Indiana University, Bloomington, Indiana , United States of America

[4] Department of Information and Library Science, Luddy School of Informatics, Computing, and Engineering, Indiana University, Bloomington, Indiana, United States of America

* Corresponding author

Email: abueckle@iu.edu (AB)

## Abstract

Working with organs and extracted tissue blocks is an essential task in many medical surgery and anatomy environments. In order to prepare specimens from human donors for further analysis, wet-bench workers must properly dissect human tissue and collect metadata for downstream analysis, including information about the spatial origin of tissue. The Registration User Interface (RUI) was developed to allow stakeholders in the Human Biomolecular Atlas

Program (HuBMAP) to register tissue blocks—i.e., to record the size, position, and orientation of human tissue data with regard to reference organs. The RUI has been used by tissue mapping centers across the HuBMAP consortium to register a total of 45 kidney, spleen, and colon tissue blocks, with planned support for 17 organs in the near future. In this paper, we compare three setups for registering one 3D tissue block object to another 3D reference organ (target) object. The first setup is a *2D Desktop* implementation featuring a traditional screen, mouse, and keyboard interface. The remaining setups are both virtual reality (VR) versions of the RUI: *VR Tabletop*, where users sit at a physical desk which is replicated in virtual space; *VR Standup*, where users stand upright while performing their tasks. All three setups were implemented using the Unity game engine. We then ran a user study for these three setups involving 42 human subjects completing 14 increasingly difficult and then 30 identical tasks in sequence and reporting position accuracy, rotation accuracy, completion time, and satisfaction. All study materials were made available in support of future study replication, alongside videos documenting our setups. We found that while VR Tabletop and VR Standup users are about **three times as fast** and about **a third more accurate** in terms of **rotation** than 2D Desktop users (for the sequence of 30 identical tasks), there are no significant differences between the three setups for **position accuracy** when normalized by the height of the virtual kidney across setups. When extrapolating from the 2D Desktop setup with a 113-mm-tall kidney, the absolute performance values for the 2D Desktop version (**22.6 seconds** per task, **5.88 degrees rotation**, and **1.32 mm position** accuracy after **8.3 tasks** in the series of 30 identical tasks) confirm that the 2D Desktop interface is well-suited for allowing users in HuBMAP to register tissue blocks at a speed and accuracy that meets the needs of experts performing tissue dissection. In addition, the 2D Desktop setup is cheaper, easier to learn, and more practical for wet-bench environments than the VR setups.

# 1. Introduction

The human body consists of trillions of cells. Understanding what cells exist in which anatomical structures and spatial contexts is essential for developing novel approaches to curing diseases. HuBMAP is a research effort carried out by hundreds of researchers in several dozen institutions in the U.S. and abroad [1]. The goal of the multi-year project is to create a reference atlas of the healthy human body at single-cell resolution, capturing spatial information about cells and tissues in unprecedented detail. In order to facilitate HuBMAP's ambitious mission, different tools are being developed. This paper presents a 3D object manipulation user interface, called the Registration User Interface or (RUI), developed to support tissue registration performed by tissue mapping centers (TMCs), as well as Transformative Technology Development (TTD) and Rapid Technology Implementation (RTI) teams. In the remainder of this section, we review typical approaches to registering tissue data together with registration accuracy typically achieved. We then derive a list of requirements for our approach to tissue registration and discuss research questions and hypotheses. The purpose of this paper is threefold: First, we compare three rather different possible setups of the RUI in a user study using telemetry, analysis, and visualization. As will be explained later in this section, this type of tissue registration is typically done via photography and note-taking. Second, in our quest to validate the 2D Desktop setup, we collect data on position accuracy, rotation accuracy, completion time, and satisfaction for the 2D Desktop implementation. Third, by virtue of the Plateau phase, we aim to understand how long it takes users to achieve a level of maximum output given a usage scenario that mimics real-world tissue registration needs, even though the three solutions presented in our paper are neither entirely application-driven nor developed with the goal of advancing 3D manipulation as a whole.

## 1.1 Tissue registration procedure and prior work

Developing a human reference atlas at single-cell resolution requires recording the size, position, and rotation of tissue extracted from living or post-mortem patients—before the tissue is processed for spatially explicit analysis. **Fig 1**A shows a photo of a typical setup: a kidney was butterflied and placed on a dissecting board to capture its size and shape, as well as the size, position, and rotation of a tissue block (outlined in pink) extracted from it. Commonly, a computer is close to the dissection work area so data can be entered and uploaded.

The documentation of extraction sites is non-trivial as different donors might have organs of different sizes and the number and shape of anatomical structures (e.g., the number of renal pyramids per kidney) might differ across individuals. It is common to use exemplary organs derived from an individual donor's data as a reference. An example is the male left kidney derived from the Visible Human (VH) dataset [2] published by the National Library of Medicine (NLM). This 3D model is about 100 mm high (see green y-axis), 60 mm wide (red x-axis), and 40 mm deep (blue z-axis)—see **Fig 1**, B and C.

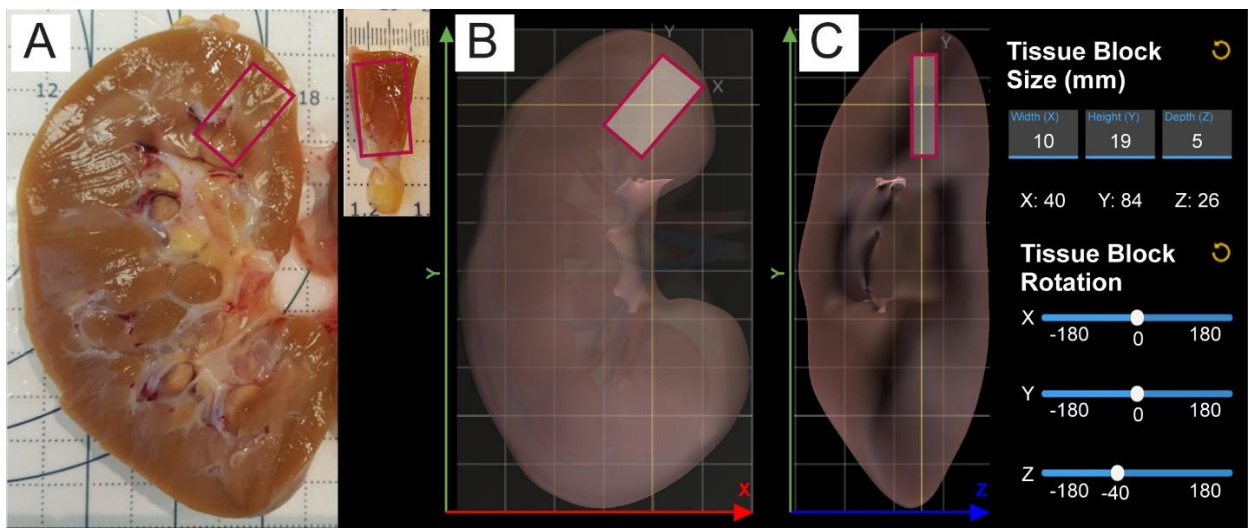

**Fig 1. Physical vs. virtual tissue registration.**
**(**A) Bisected kidney on a dissecting board. Pink outlines indicate where the tissue block highlighted pink (shown in top right) will be extracted. (B**)** RUI with reference kidney of about the same size in x-y view. (C**)** RUI in z-y view with user interface that supports entry of tissue block size in mm, review of x, y, z position values, and change of tissue block rotation in 3D.

Shown in **Fig 1**C is the interface used for entering 'Tissue Block Size' in mm and for rotating the tissue block (via sliders). Position can be adjusted by dragging the tissue block to the correct location within the 3D reference organ. An x-y view and a y-z view can be selected to check and correct the tissue position. The x, y, and z positions are also displayed to the user. Size, position, and rotation values can be reset by clicking on the corresponding circular arrow buttons.

Different procedures exist to capture relevant information; resulting data is submitted to diverse clinical record-keeping systems with different metadata schemas. Different organs—e.g., lung [3], breast [4], thymus [5], and pancreas [6]—have rather different needs and are subject to many standard operating procedures (SOPs) and checklists [7, 8]. A closer look at these SOPs and checklists developed     for different organs by different authors reveals the lack of common procedures and documentation standards. More importantly for HuBMAP, existing data captures only **partial or inconsistent spatial information** (i.e., the level of detail at which this information is captured varies across protocols).

### 1.1.1 Partial and inconsistent spatial information

Pathologists and other wet-bench workers typically use SOPs—written as protocols [9, 10] and published on protocols.io—to ensure reproducibility, establish relevant terminology, and share otherwise disparate materials and instructions in a consistent framework. Most importantly for our research, they use SOPs to capture specific workflows such as extracting tissue blocks from organs [11], tissue preservation through freezing [10, 12], or preparing specimens for further analysis (9). Some of these protocols require the lab worker to capture the spatial origin of tissue in reference to an organ and/or its dimensions: e.g., some SOPs involve pictures of dissected organs or tissue blocks on dissecting boards with markings for length and diameter units [11-13] (see **Fig 1**A), occasionally at different stages of the dissection process [14]. The scale of the marker positions and the reported data is in the millimeter range. When using dissecting boards with markings is not feasible, some protocols supply abstract illustrations to show the extraction sites of tissue [15]. The quality and purpose of these pictures vary; many are ad hoc, with inconsistent lighting and varying quality, or no pictures at all [16]. In some cases, the authors provide no exemplary pictures but give a verbal description of how the donor organ has to be aligned and dissected [17]. This causes many of the existing protocols to capture only **partial and inconsistent spatial information**. Manual annotations in these pictures offer a small amount of orientation with regard to the spatial provenance of the tissue block, but this kind of documentation lacks detail and reproducibility across teams and organs. Further, inferring the correct dimensions of a tissue block from a photo can be challenging, depending on the distance between the side of the tissue facing the viewer and the cutting board.

### 1.1.2 Limited computability of photos

A second issue with the current record-keeping practices for spatial origins of tissue blocks is that **images** of extracted tissue and/or organs, if present at all, **are not computable**. To be of value for the HuBMAP atlas, tissue spatial data must be provided in a format that is uniform across organs and can be used to correctly determine the size, position, and rotation of tissue blocks in relation to a 3D reference body. Images with spatial annotations do not support this, and advanced techniques such as computer vision algorithms cannot be trained and used due to the quality and limited quantity of existing images. While photos provide an efficient way of archiving general spatial information in the context of individual labs, they do not provide the precision and standardization required for reference atlas design. To overcome this limitation, we implemented an online service that lets subject matter experts (SMEs) size and register 3D tissue blocks within 3D reference organs to generate unified data across tissue types.

### 1.1.3 Challenges of 3D manipulation

The Registration User Interface (RUI) was developed to **address practical concerns** regarding tissue registration. However, there are several known challenges when manipulating 3D objects in 3D. We assume the SME is an able-bodied individual with two hands and a basic understanding of how to use photography equipment and a paper or digital documentation sheet. The SME places the tissue on the dissecting board, aligns it with the provided grid system, takes photos, and writes down annotations.

In the proposed RUI, there are various cognitive challenges as 3D manipulation is non-trivial. In our review of prior work, we focus on two methods to enable a user to manipulate a virtual object in 3D space: widgets and extended input devices. The de-facto standard in many 3D modeling applications is the use of a mouse and color-coded virtual **widgets** attached to the object, as discussed in Maya [18] and Blender [19]. These widgets allow the user to perform position, rotation, and scaling operations. Schmidt, Singh and Balakrishnan [20] proposed a user-input-based extension to the traditional widget system. In a pilot study with assembly tasks, they found that the most experienced participants users needed twice as long when using their system than when using the traditional version while users with average familiarity required roughly the same amount of time with both methods (under instruction), pointing at the difficulty of overcoming existing usage patterns and expectations for 3D manipulation tool by experts. A similar framework was proposed in 1995 by Bukowski and Séquin [21], who prototyped an interaction language for "pseudo-physical behavior" from the user, where the 2D motion of the mouse cursor is extrapolated into 3D motion in the virtual environment. These "object associations" employed properties such as gravity to make the alignment of 3D objects more intuitive for the user, .

In addition to using widgets to perform 3D manipulation, there exists a body of literature about **input devices extending the standard computer mouse** for these tasks. Balakrishnan et al. [22] developed the "Rockin'Mouse", a four-degree-of-freedom (DoF) device that allowed users to control position and rotation without having to switch between modes. Their pilot study found that users were able to complete a set of block-matching tasks 30% faster with a Rockin'Mouse than with a regular mouse. In order to explore the design space of a multi-touch mouse, Villar et al. [23] presented a series of five prototypes using different touch input layouts. They found that ergonomics and form-factor were important design aspects for user satisfaction, although their study was aimed at gathering qualitative results rather than quantitative performance measures. For their GlobeMouse and GlobeFish, Froehlich et al. [24] separated position and rotation manipulation using a trackball (rotation) connected to an inner and outer frame (position). When tested against a commercial option in a study, they found that the completion times for their devices were significantly faster than for the commercial SpaceMouse, although they found a similarly strong learning effect for the three devices tested over the course of two sets of four tasks per device (with training sessions before each task). A commercial approach to the extended mouse is the aforementioned SpaceMouse [25], a six-DoF device that lets a user position and rotate a 3D object along and around all three axes at the same time using a self-resetting internal mechanism when no user input is given. A major issue for this advanced, modified hardware is the steep learning curve, making its widespread deployment and adoption unlikely.

While the aforementioned projects feature variations on the widely used mouse input device, recent efforts have focused on alternative input devices. Soh et al. [26] developed a simple hand gesture interface for the Microsoft Kinect to enable translation of rotation of 3D objects. Similarly, Lee et al. [27] used a webcam and a projector to transform a piece of cardboard into a movable, handheld 3D device that lets users rotate the projected 3D object. In a more recent paper, Mendes et al. [28] used the HTC Vive and Unity to design a system for custom translation and manipulation axes (MAiOR). In a user study comparing MAiOR to a regular six-DoF approach without separation of manipulation and rotation as well as a system with virtual widgets, they found that the approach with traditional widgets achieved the highest overall success rate but came at the cost of higher completion times with increasing task difficulty and confirmed that mid-air manipulations with VR controllers lack precision.

### 1.1.4 Overcoming the challenges of 3D manipulation

Building on and extending this prior work, the HuBMAP RUI aims to support scalable and computable tissue registration and data management. It lets experts use a nearby computer to digitally capture the size, position, and rotation of tissue blocks in relation to a reference organ, together with important metadata such as name, tissue ID, date, and time.

This paper presents the results of user studies that aim to determine and compare registration accuracy and speed for different user interfaces. Specifically, we compare the 2D desktop setup with two VR setups—using a sitting and a standing setup. All three user interfaces support the general registration task shown in **Fig 2** and detailed in Section 2.2. In all three setups, the subject sees a reference organ kidney (modeled after the aforementioned VH kidney, see section 1.1) with a virtual purple target block on the left and a white tissue block on the right that needs to be matched in position and rotation with the target block. In all cases, the sizes of the tissue block and the target block are identical, which resembles the real-world scenario in which a tissue block has just been extracted from an organ. We implemented the tissue block as a cuboid as this primitive shape mimics the approximate dimensions of real-world tissue samples, and because testing multiple shapes would have been out of the scope of this study.

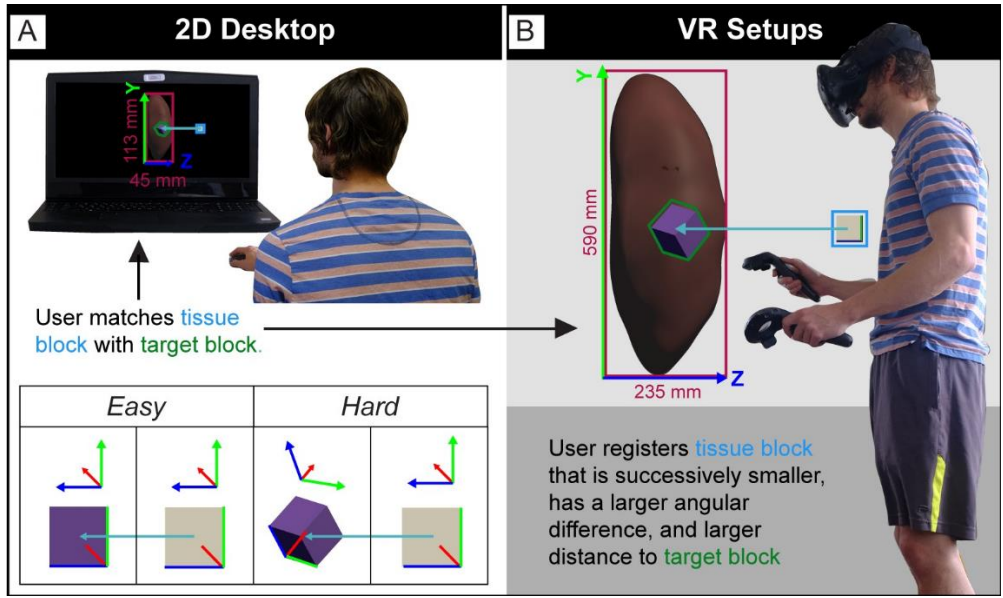

**Fig 2. The task setup in our user study.**
Reference organ with target block indicated (purple) and tissue block (white) to be registered into the target block. The light blue arrow indicates block centroid (mid-point) distance. Task difficulty increases as the tissue blocks get smaller, block rotation increases, and distance between the blocks increases. (A) 2D Desktop setup. (B) The two VR setups.

The reference organ in our study appears in different sizes in the 2D Desktop setup and the two VR setups (see **Fig 2**). The kidney is 113 mm tall on the screen in the 2D Desktop condition and 590 mm tall in VR. We chose these different sizes to make use of the ability in VR to interact with objects that would not normally fit on a regular laptop screen. We elaborate on this (and its implications for data analysis) when presenting the results in Section 3. Three color-coded coordinate axes (implemented as long, thin cylinders running along the edges of each cube, conjoint at one of its corners) are used to indicate tissue and target block rotation. They are colored red, green, and blue for x-, y-, and z-axis, respectively. As we present the users with increasingly difficult registration tasks (see **Fig 4**), subjects must adjust not only the position but also the rotation of the tissue block to match the position and rotation of the target block. The focus on these adjustments mirrors the real-world need for these 3D manipulations to capture the spatial provenance of tissue blocks with regard to a reference organ in the RUI.

## 1.2 Requirements for registration user interface

Informal interviews and registration tests with pilot study participants revealed various requirements for the RUI. Requirements can be grouped into five categories and are discussed subsequently.

1. **Metadata Entry**: The RUI must support entry of data such as user name, organ name, tissue block size, and date and time of registration. This metadata must then be sent to a database for ingestion and usage in the HuBMAP data infrastructure and portal.
2. **Accuracy**: The RUI must support gross-anatomical-tissue registration at about **one mm** for position and about **20 degrees** for rotation accuracy.
3. **Training and Completion time**: The RUI should not require more than **five minutes** to learn, and each tissue registration should not take more than **one minute** to complete.
4. **Satisfaction**: The RUI should be easy to use, and subjects should feel a sense of accomplishment after they perform the registration task.
5. **Deployment**: The RUI should be usable on a computer in a lab, ideally right after tissue has been extracted. A typical lab computer uses a Windows or Mac operating system and runs Chrome, Firefox, or other web browsers. A typical window size is 1920 x 1080 (full HD) or 3840 x 2160 (4K) pixels at 72 DPI.

## 1.3 Research questions and hypotheses

Given the overall domain task and the requirements stated in Section 1.2, there are six research questions (RQ) this study aims to answer. We also present associated hypotheses (H) here:

**RQ1**: What position and rotation accuracy and completion time can be achieved with the three different RUI setups?

**H1a**: Users in VR Tabletop and VR Standup achieve significantly higher position accuracy than users in 2D Desktop.

**H1b**: Users in VR Tabletop and VR Standup achieve significantly higher rotation accuracy than users in 2D Desktop.

**H1c**: Users in VR Tabletop and VR Standup have significantly lower completion times than users in 2D Desktop.

**RQ2**: What are the error and bias, i.e., the deviations for each axis as well as the cumulative deviation (see Section 2.4.2), for position accuracy in all three dimensions?

**H2a**: We do not expect any major bias for any setup in any dimension.

**H2b**: We expect the error to be greatest for the 2D Desktop setup due to its restricted input devices and limited viewing positions.

**RQ3**: How does task complexity (e.g., smaller tissue block size or more rotation, larger distance between tissue block and target) impact accuracy and completion time?

**H3a**: More complex tasks lead to lower position accuracy for all setups.

**H3b**: More complex tasks lead to lower rotation accuracy for all setups.

**H3c**: More complex tasks lead to higher completion times for all setups.

**RQ4**: What is the maximum performance level that a user can reasonably achieve, and how many practice tasks are required before performance levels out? That is, after how many tasks do users reach a plateau when accuracy or completion time do not significantly change anymore.

**H4**: VR users need a lower number of tasks to plateau than 2D Desktop users.

**RQ5**: What is the tradeoff between accuracy and completion time? For example, if users are asked to register fast, does accuracy decrease? If users are asked to register accurately, do they need a longer time to complete the task?

**H5**: In all setups, the more time users spend on a task, the higher position and rotation accuracy they achieve.

**RQ6**: How satisfied are users with the results achieved in the three different setups?

**H6a**: Users in both VR setups are more satisfied with their performance than 2D Desktop users.

**H6b**: There is no significant difference in user satisfaction between VR Standup and VR Tabletop users.

The paper is organized as follows: In Section 2, we introduce methods, including study design, task difficulty, and performance metrics. In Section 3, we present the qualitative and quantitative results of this study before interpreting the results with regard to the requirements from Section

1.2 and the research questions and hypotheses stated in Section 1.3. In Sections 4 and 5, we discuss results and present an outlook on planned future work.

## 2. Materials and methods

This section presents the overall study design, the three different hardware/software setups, task difficulty metrics and synthetic tasks generation, as well as human performance metrics, plateau, and a satisfaction score computation. A power analysis was conducted prior to running the experiment to determine the number of subjects required to achieve significant results (see S1 Appendix).

### 2.1 Ethics statement

All subjects consented to participate in the research. When starting their participation, each subject was presented with a study information sheet (SIS), displayed to them on a computer. We then asked our participants if they had understood the SIS and informed them that by proceeding with our experiment, they were consenting to their participation. We did not collect written consent as the data was analyzed anonymously, and as this research presented no more than minimal risk of harm to subjects while involving no procedures for which written consent is normally required outside of the research context. This is in accordance with the guidance on expedited human-subject research protocols as issued by Indiana University's Institutional Review Board (IRB) for this project under IRB number 1910331127. Summary information (including the IRB application number) can be found below.

The individual shown in **Fig 2** and **Fig 3** in this manuscript has given written informed consent (as outlined in a PLOS consent form) to publish these case details.

## 2.2 Study design

We used a typical user study design featuring a study information sheet (SIS) in the beginning to get user consent, followed by a pre-questionnaire, tutorial and experiment tasks, and a post-questionnaire. All 42 subjects were run in person by one of the authors of this paper.

The main part of the experiment asked subjects to use one of three setups: 2D Desktop, VR Tabletop, or VR Standup. 2D Desktop corresponds to the RUI as deployed in HuBMAP, with a 2D interface running on a laptop, and represents a base case for 3D interaction; VR Tabletop mimics everyday work in the form of a virtual and physical desk while still allowing the user to be active in an immersive environment; VR Standup, finally, enables the subject to use their entire body for navigation around the 3D scene for full use of their motor functions and spatial abilities through the VR equipment. All three setups are shown in **Fig 3**, A-C. Users were randomly assigned to one out of these three setups. Different levels of task difficulty were used for the tutorial, Ramp-Up, and Plateau tasks (see Section 2.3). Tasks were identical for all users regardless of setups (performance metrics are detailed in Section 2.4). Users determined for themselves when a task was done and were provided the equivalent of a "Next" button in all three setups. More information can be found in videos showcasing all three setups from a user's perspective on GitHub ([https://github.com/cns-iu/rui-tissue-registration](https://github.com/cns-iu/rui-tissue-registration)). The SIS, pre-, and post-questionnaire data was presented and gathered using an online Qualtrics form. The tutorial and experiment tasks used a setup implemented in the Unity game engine [29]. The Qualtrics form, along with documentation of logged data formats and data analyses, can also be found on GitHub.

All subjects used the same Alienware 17 R4 laptop with a display diagonal of 439.42 mm (17.30 in), running Unity 2019.4 on Windows 10 with a secondary monitor attached for ease of configuring the individual steps of the experiment. The laptop had an Nvidia GTX 1070 with 32 GB RAM of memory. A 1080p webcam recorded audio and video. For the VR setups, we used a 2016 HTC Vive with two Vive controllers. We ran the application for all three setups straight out of Unity. Data was collected using a custom C# script, writing data to a CSV file at a frequency of 10 Hz every time the user pressed a button.

The research facilitator could observe the subjects' viewpoint on the laptop display, which was recorded with a screen-capturing software. We conducted the study in a collaborative space in a public university building and took precautions to preserve our subjects' safety. The usable space for VR Standup and VR Tabletop users was around 10 x 10 ft (3 x 3 m). 2D Desktop users sat at a 4 x 4 ft table (1.2 x 1.2 m).

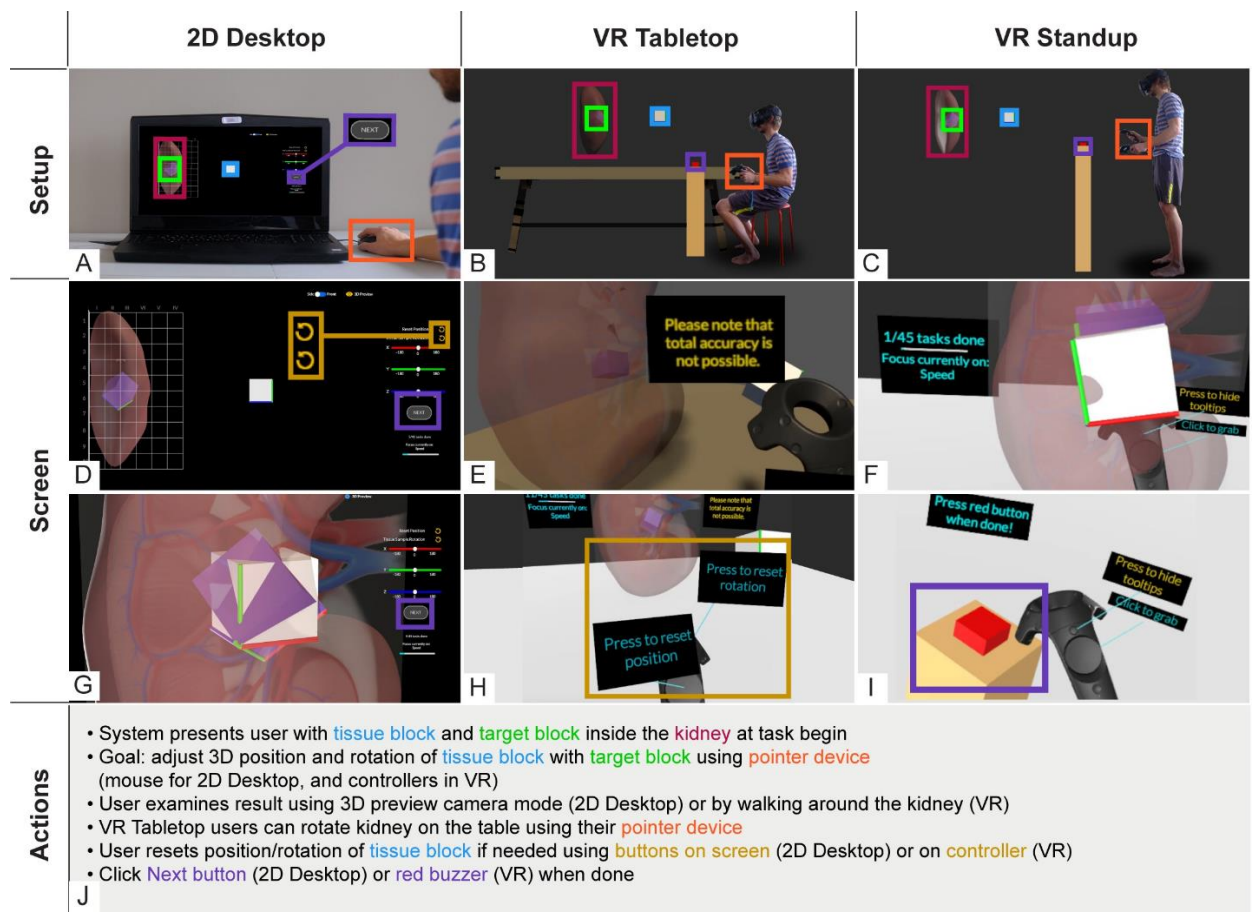

**Fig 3. Setup, screen, and actions for 2D Desktop, VR Tabletop, and VR Standup.**
(A-C) Three RUI setups with a human subject. (D-I) screenshots of the user interface. (J) Required actions. The tissue block is outlined in blue, the target block in green, and the kidney—providing context and domain relevance—in pink. Tasks are submitted by selecting the purple NEXT/red button. The user could reset the position or rotation of the tissue block by pressing the corresponding yellow-brown virtual (2D Desktop) and physical buttons (VR).

The three setups support nearly identical functionality as detailed in section 2.2.1 below.

Our study design was inspired by a variety of prior work. Batch et al. [30], in their paper on evaluating a VR visualization framework for economists, lay out a telemetry setup using the same hardware used in our study, and after which we modeled our data collection approach (focusing their analysis, among other metrics, on completion time and positions of HMD and controllers). Regarding the comparison of different setups or modalities, we reviewed related work by Millais, Jones and Kelly [31] on comparing user performance and satisfaction when exploring scatter graphs and parallel coordinate plots in Desktop vs. VR environments using a Google Daydream VR HMD. Further, Cordeil et al. [32] ran a user study with two cohorts (Oculus Rift VR HMD and CAVE-style system [33]) where they asked participants to perform collaborative analysis of an abstract network visualization. Similarly, subjects in a study by Prabhat et al. [34] were measured for their performance and subjective evaluation of three display conditions while analyzing biological data (Desktop, Fishtank display, CAVE [33]). Our idea to compare not only Desktop vs. VR setups but also different VR implementations with each other was partially inspired by a study by Kwon et al. [35], who conducted a user study on the impact of different layouts, rendering methods, and interactions for graph exploration while measuring task completion time, accuracy, and number of interactions on an Oculus Rift DK2. Finally, Cikajlo and Potisk [36] researched the influence of VR vs. Desktop interventions on performance in patients with Parkinson's disease when placing cubes, using the Oculus Rift CV1.

### 2.2.1 The three setups

As described earlier, in both VR setups, the kidney was around 590 mm tall. In 2D Desktop, on the 1080p screen, the kidney appeared at a height of 113 mm. We chose these different sizes to make use of the ability in VR to display and interact with 3D objects in a much larger size than it would be possible on a standard laptop display. **Fig 3**, A-C, illustrates these differences with a human subject for scale.

2D Desktop

The screen in the 2D Desktop setup (see **Fig 3**A, D, G) consisted of a 3D work area covering most of the screen and featured a transparent model of a human kidney with an inlaid 2D image showing a schematic drawing of the vasculature with a 100 x 60 x 60 mm grid wrapping around the kidney. We added a progress bar on the bottom right with a text field indicating the number of completed tasks.

In terms of controls, the user could move the tissue block by clicking and dragging it with the left mouse button; they could also rotate it around each axis with the three sliders on the right side of the screen. Both position and rotation could be reset via button clicks.

The setup had two cameras the user could choose between: an orthographic camera (always aligned with either the x- or the z-axis) and a perspective ("preview") camera. The user could switch between the two by clicking the yellow eye icon on the center right, top edge of the screen. This preview camera could be rotated with relative freedom by clicking and dragging the left mouse button (see **Fig 3**G). The main camera, however, was less movable. The toggle switch in the top center allowed a movement of 90 degrees around the kidney's upward-facing axis, allowing the user to go back and forth between two predefined viewpoints with a smooth, animated transition. The usage of an orthographic camera is common in 3D modeling software as it makes the alignment of objects easier by taking away one dimension. The user could proceed to the next task by clicking a "Next" button.

### VR Tabletop

In the VR Tabletop setup (see **Fig 3**B, E, H) the user was presented with a 3D model of a kidney and the same tissue block and target block as 2D Desktop users. As the name implies, the user sat at a table, both in the physical and virtual world. This allowed us to test whether simulating a physical work desk environment helped with the 3D alignment. The functionality provided through UI buttons and the mouse in the Desktop setup was implemented using a VR headset and VR controllers as pointer devices. The trigger button on the right hand allowed the user to grab and move the tissue block. Pressing the left touchpad or menu triggered a reset animation for the position or rotation of the tissue block, respectively. By default, tooltips were displayed atop the controllers, which could be turned off by pressing the right menu button. The user could rotate the kidney around its y-axis with the touchpad on their left hand and move the tissue slice with their right hand. The ability to turn the kidney was unique to VR Tabletop. The user could proceed to the next task by touching a virtual red buzzer on a stand at a height of around three ft (90 cm) above floor level.

VR Standup

This setup (**Fig 3**C, F, I) was similar to VR Tabletop; however, users stood in front of a reference kidney, able to explore it from 360 degrees while being assisted by the research facilitator for physical safety. The user could not rotate the kidney in this setup but was able to walk around the kidney to see it from different viewpoints. Otherwise, the implementation was identical, with VR controllers as pointer devices and a virtual red buzzer to proceed to the next task.

### 2.2.2 Pre-questionnaire

After being welcomed to the experiment, subjects were asked to fill out a pre-questionnaire using an online survey software running on the same laptop we used for the actual tasks. This pre-questionnaire inquired about the subjects' prior experience with virtual reality and 3D video games and about their familiarity with different types of data visualizations such as graphs, charts, tables, maps, and networks. Additionally, we asked our subjects to disclose demographic information such as native language, job title, age, and gender. Items of specific interest for our user study were also whether users were right-handed or left-handed, their height, and whether they had a vision impairment. The complete questionnaire is available at https://github.com/cns-iu/rui-tissue-registration.

### 2.2.3 Tutorial task

After answering the pre-questionnaire, the subject was either presented with the experiment application in Unity (2D Desktop) or they donned the VR gear and got into position (VR Tabletop and VR Standup). They then listened to an approximately three-minute audio tutorial explaining the elements in the scene (kidney, tissue block, target block), what the tasks entailed, how to move and rotate the tissue block in a given setup, how to reset the position and location of the tissue block if needed, and how to submit task results and get a new task. The prerecorded audio ensured the same delivery of the content to all subjects. While the audio was playing, the subject was encouraged to practice moving and rotating the first tissue block, and to explore the 2D screen or 3D scene in front of them. We encouraged subjects in VR Tabletop and VR Standup to quit the experiment should they feel nauseous. The research facilitator monitored subjects at all times to ensure their physical safety. Having a facilitator present also helped many first-time VR subjects to correctly strap on the headset and move between VR and physical world without damage to the equipment or themselves.

### 2.2.4 Ramp-up tasks

Following the tutorial task, in the Ramp-Up phase, we asked subjects to solve 14 increasingly difficult tasks over time (see explanation of task difficulty in Section 2.3). As the task numbers increased, the size of the blocks to be placed became smaller and the rotational differences and distance between tissue and target block increased. After a pilot study with eight subjects, we decided to use 14 tasks to cover major difficulty levels while allowing every subject to finish the entire experiment in 60 minutes.

In the pilot study, some subjects spent unusually long times in the VR setups to achieve near perfect accuracy. To avoid this, we added three interventions: First, during the tutorial, we mentioned that 100% accuracy was not possible and asked subjects to use their best judgement when determining whether they were done with a task. Second, in the VR setups, we added a constantly visible text box next to the kidney with a reminder that 100% accuracy was not possible (see **Fig 3**E). Third, we gave subjects alternating audio prompts for odd tasks (focus on speed) and even tasks (focus on accuracy) in all three setups—this also let us explore the influence of task complexity on accuracy and completion time (RQ3, see Section 3.4), and tradeoffs in speed versus accuracy (RQ5, see Section 3.5).

### 2.2.5 Plateau tasks

Interested in understanding the number of tasks it takes before a user achieves their maximum performance in terms of accuracy and completion time, we asked subjects to register 30 blocks of identical size as fast as possible during the Plateau phase (RQ4, with results in Sections 3.2 and 3.3). The Plateau phase followed immediately after the last task of the Ramp-Up phase. We determined this number of tasks in pilot studies where we aimed to achieve a balance between subject exhaustion, total participation time, and detectability of a plateau. Given that subjects in the Desktop setup spent around three times longer on tasks as subjects in the VR setups (see Section 3.3), the number of tasks in the Plateau phase had to be high enough for us to detect a performance plateau while ensuring a timely finish of the experiment. A more complete description of Plateau phase task difficulty can be found in Section 2.3 and **Fig 4**.

### 2.2.6 Post-questionnaire

After finishing the registration tasks part of the study, the Unity application was closed, and each subject (now out of VR if part of VR Tabletop or VR Standup) completed a post-questionnaire about their experience. We included this post-assessment to learn how much users liked the registration interface, to determine what they would improve, and to compare user satisfaction across setups. Satisfaction score compilation is detailed in Section 2.4.6.

## 2.3 Task difficulty and stimuli generation

Task difficulty in the Ramp-Up phase of our study is a combination of the distance between tissue block and target block, the size of both blocks, and the angular difference between both blocks (see **Fig 4**). During the Ramp-Up phase, distance, angular difference, and size were continuously increased until the distance was 200% of the kidney height, the angular difference was 180 degrees, and the side length of the two blocks were only 5% of the kidney height. To generate stimuli used in the Ramp-Up phase, we used Lerp(), a native Unity method for linear interpolation (see **Equation 1**) to interpolate between start and end values. To increase angular difference over time, we used Slerp(), a different implementation of the aforementioned Lerp() function (for rotations), to gradually rotate the target block towards an end rotation of 0, 270, and 180 (around the x-, y-, and z-axis) using linear interpolation.

$$Lerp = \begin{cases} a, & \text{if } t \leq 0 \\ b, & \text{if } t \geq 1 \\ a + (b - a) * t, & \text{if } 0 < t < 1 \end{cases}$$

**Equation 1. Formula to compute task difficulty.**

Distance and angular difference were smallest in the tutorial at 30% of the kidney height and 0 degrees rotational difference. Similarly, the length of each edge in both blocks was 20% of the kidney height, initially. Finally, in the Plateau phase, these values were consistent, with the distance and size values at around the same level of difficulty as the average Ramp-Up task (115% the kidney height for distance and 12.5% of the kidney height for size) but at maximum angular difference. Note that while we only show one Plateau task in **Fig 4**, there were 30 identical ones. **Fig 4** also shows the tutorial (simplest), 14 increasingly difficult Ramp-Up tasks, and the 30 Plateau tasks (all of same task difficulty). Details on sizes, rotations, and distances used for the tasks are provided together with information on audio prompts.

| | Distance | Angular difference | Size difference | Number of tasks | Prompt |
|---|---|---|---|---|---|
| Tutorial | *Smallest* | | *Biggest* | | |
| | 0.3 * kidneyHeight + offset | 0 degrees + offset | 0.2 * kidney Height | 1 | Audio explanation of interface |
| Ramp-Up | *Increasing* | | *Decreasing* | | |
| | start: 0.3 * kidney Height + offset end: 2 * kidneyHeight | start: 0 degrees + offset end: 180 degrees | start: 0.2 * kidney Height + offset end: 0.05 * kidney Height | 14 | Odd: speed Even: accuracy |
| Plateau | *Consistent* | | *Consistent* | | |
| | 1.15 * kidneyHeight | 180 degrees | 0.125 * kidneyHeight | 30 | Speed |

**Fig 4. Task setup and levels of difficulty used in this study**.
Distance, angular difference, size difference, number of tasks, and prompt type for the one Tutorial, 14 Ramp-Up, and 30 Plateau tasks. The offset (computed via **Equation 1**) is a value that is added to gradually increase the distance and angular difference between the two blocks, and that is used to gradually decrease the size of the two blocks. Note that due to the layout of this figure, only 13 out of the 14 Ramp-Up tasks are illustrated on the left.

The computation of Lerp() requires three values, where *a* is the start value (easiest), *b* is the end value (hardest), and *t* is an interpolation value between 0 and 1. For every task, *t* is computed by dividing the current task number by the total amount of Ramp-Up tasks (14). At task number 0 (i.e., the tutorial task), *t* evaluates to 0, which causes the function to return the start value. At task number 14 (the last Ramp-Up task), *t* evaluates to 1, prompting the function to return the end value. For any task in between, the function returns the start value with an offset value that increases over time. As input, Lerp() uses 3D vectors while Slerp() uses 3D rotations.

To ensure that task difficulty is identical across setups, we normalized each of the difficulty parameters (distance, rotational difference, size) by the height of the kidney in each condition.

**Fig 4** shows the tissue blocks and target blocks used throughout the experiment alongside the kidney for scale. The other columns indicate the different values for distance, angular difference, size, number of tasks, and audio prompts.

## 2.4 Performance metrics, plateau, and satisfaction score

To analyze survey and task data, we defined three performance metrics (position accuracy, rotation accuracy, and completion time) as well as a satisfaction score.

### 2.4.1 3D position accuracy

To answer RQ1 and RQ2, we needed to assess the position accuracy for each subject. Position accuracy equals the distance of the centroids of the tissue block and the target block, see light blue arrow in **Fig 2**. We compute the distance at run time using `Vector3.Distance()`, a static method in Unity that returns the distance between two points in 3D space. The position of both blocks and the centroid distance was collected at 10 Hz (i.e., 10 times each second).

To make use of the various possibilities for scaling in VR, the kidney was displayed in different heights across setups (but always with the same width-to-height-to-depth ratio). Measured from the lowest to the topmost vertex, the kidney in the two VR setups was 0.59 Unity scene units tall. In VR, scene units correspond to physical meters, so the kidney appeared at a height of 590 mm. Similarly, in the 2D Desktop setup, the kidney appeared at a height of 113 mm on the laptop display (see **Fig 2Error! Reference source not found.**). In order to compare position accuracy results between 2D Desktop and the VR setups, we normalized these values by dividing them by the height in which the kidney appeared to the user. When discussing the results in Section 3.2, we append the subscript "$_{norm}$" to denote normalized position accuracy values.

### 2.4.2 Bias and error

We also recorded raw position data for both blocks to compute *bias* and *error* for each tissue block placement (see Section 3.2). We define error as the median distance from every placed tissue block from the target block. This can be computed for all three dimensions, enabling us to describe position accuracy in a higher precision than just relying on a one-dimensional distance value. Bias, on the other hand, is the three-dimensional Euclidean distance $d(p,q)$ with the Cartesian coordinates for $p$ being the target centroid position normalized to 0 and the coordinates of $q$ being the median errors in the x, y and z-dimensions. This can be computed for all three dimensions, enabling us to describe position accuracy more precisely than just relying on a one-dimensional distance value.

### 2.4.3 3D rotation accuracy

Rotation accuracy equals the angular difference between the two tissue blocks at task submission (see **Fig 2**). For ease of analysis, it was reduced to an individual number between 0 (exact same rotation) and 180 (diametrically opposite rotation). We used Unity's built-in Quaternion.Angle() function to compute this angle. Angle() takes two orientations, each consisting of three angles, expressed either as Euler angles or Quaternions, and returns a single float value between 0 and 180.

This means that several combinations of different rotations between tissue block and target block could yield the same angular difference. In order to preserve as much detail about the subject's action as possible, equivalent to the position, we logged the rotation of both blocks throughout the experiment as well. This allowed us to analyze the angular difference for all three axes (see Section 3.2).

### 2.4.4 Completion time

Completion time refers to the amount of time between the submission of a task and the submission of the previous task. Completion time is measured in seconds.

### 2.4.5 Performance plateau

During the Plateau phase (see Section 2.2.5 and **Fig 4**), subjects performed 30 identical tasks, providing a unique opportunity to identify if and when a subject achieved a performance plateau. A plateau of a performance variable (task completion time, centroid accuracy, or rotation accuracy) was reached when the deviation of the performance variable did not exceed the mean performance of the subject until the end of the Plateau phase. As mean performance, we consider the average performance in a moving window of 20 tasks of the subject to reduce the influence of possible performance outliers. This width of the moving window supplied a stable mean by considering a certain inertia in performance improvement without including at all times the extreme values that can often be found towards the beginning and the end of the Plateau phase. For each subject, we analyzed after which task the performance stabilized by iterating through a recursive process, in which the relative deviation of the last task of the Plateau phase was calculated. If it did not exceed one (thus if the deviation of the performance variable in this task is not higher than its mean), we iterated this calculation for the previous task until we arrived at a task where the relative deviation was larger than 1. We considered all tasks after this (until the last task of the phase) to be on a performance plateau. If a subject reached a performance plateau, we took the average performance (for example, mean completion time per task) of all the tasks that were completed after reaching this plateau.

### 2.4.6 Satisfaction

To assess user satisfaction, we included a corresponding item in the post-questionnaire via a five-point Likert scale from one (not at all satisfied) to five (very much satisfied), with three being a neutral value, and we report results aggregated by setup. This pertains to RQ6 (with results presented in Section 3.6).

# 3. Results

This section presents subject demographics, performance and satisfaction for all three setups, and a comparison of results plus discussion of requirements and research questions presented in Sections 1.2 and 1.3.

## 3.1 Demographics

We solicited 43 subjects in a large, Midwestern public university town via email, word-of-mouth, and social media for in-person user study appointments between 30 and 60 mins. We had to drop one subject from the analysis for not meeting the age requirement for participation (they were not 18 yet by the time of participation in the study), leaving us with 42 subjects. Subjects spent an average of 43 minutes with the experiment, including pre- and post-questionnaire. None of the subjects had medical or anatomical expertise.

The gender split in our experiment was almost exactly 50/50, with 20 female and 21 male subjects and one subject preferring not to specify gender. In terms of age, 10 were between 18 and 20 years old, 29 were between 21 and 30, one between 31 and 40, and two between 51 and 60. There were 34 English, four Chinese, two Bengali, one Russian, and one Spanish native speaker. All subjects except one were right-handed. In terms of vision impairments, 20 indicated near-sightedness, four far-sightedness, three preferred not to answer, two reported astigmatism, one reported to be both far- and near-sighted, and one presbyopia. 11 subjects reported perfect vision. All subjects were allowed to wear glasses during the experiment.

## 3.2 Accuracy

To answer RQ1 and RQ2, we analyzed the data for differences in position and rotation accuracy in the tissue block placements between the three setups for the Plateau phase. Results are plotted in **Fig 5**.

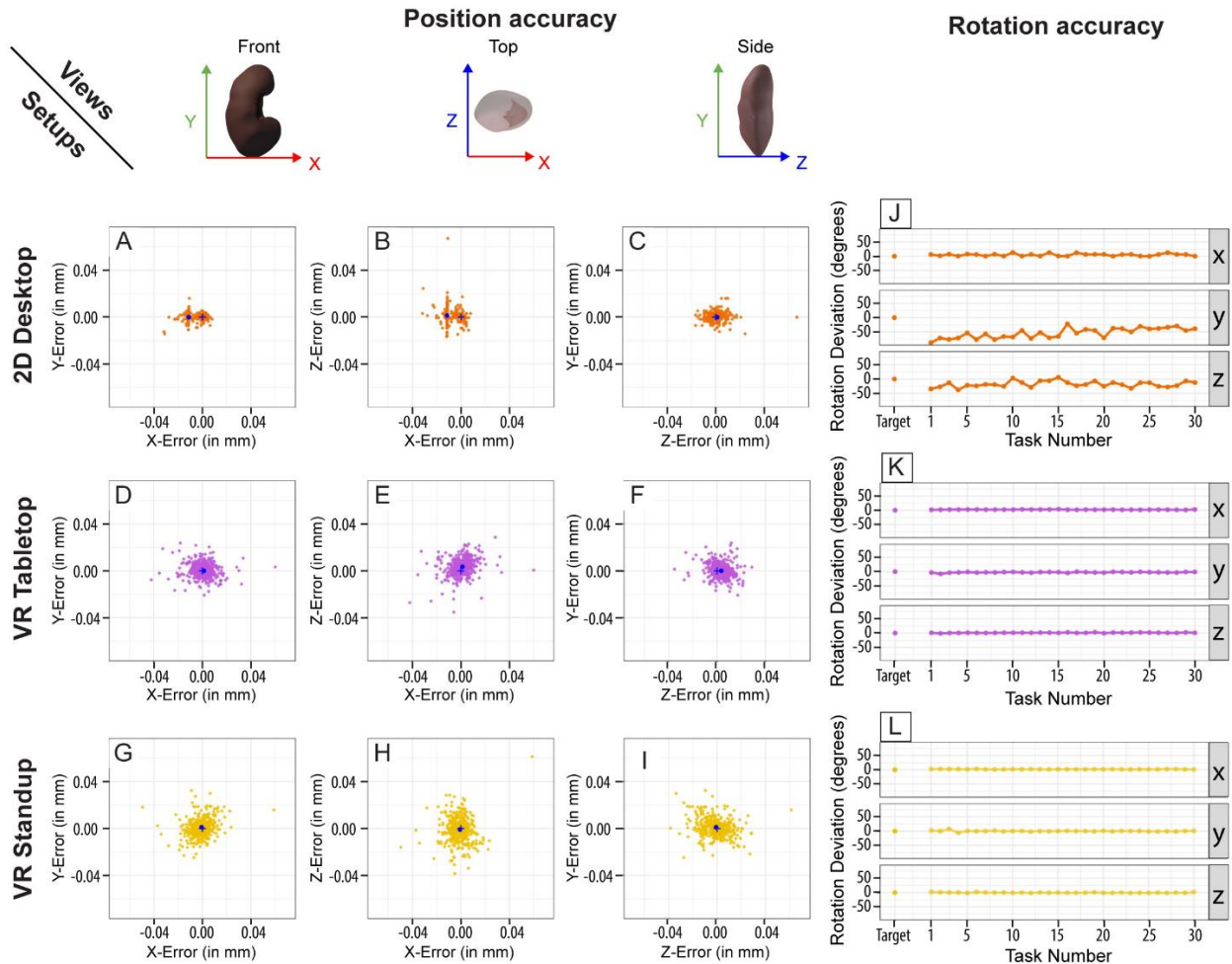

**Fig 5. Graphs for position and rotation accuracy.**
(A-I) Scatter graphs showing the error for position accuracy (in mm), normalized by kidney height, during the Plateau phase. Each dot represents one of the 30 tissue block placements. The blue cross at the origin of each scatter graph shows the location of the target block. The blue dot shows the average of all centroids (bias). (J-L) Line graphs with rotation accuracy for each axis (x, y, z).

When comparing **distance** between the centroids of the blocks for each user after they had reached their position accuracy plateau during the Plateau phase, we found **no significant**

**difference** between 2D Desktop, VR Tabletop, and VR Standup subjects when normalized by the height of the kidney (2D Desktop$_{norm}$ = 0.0118, VR Tabletop$_{norm}$ = 0.0114, VR Standup$_{norm}$ = 0.0125), prompting us to **reject H1a**. Investigating the error for each axis, we found one error that stood out: 2D Desktop subjects tended to place the tissue blocks towards the negative space of the x-axis (median x-error$_{norm}$ = -0.01138; see **Fig 5**, A and B). This error for the 2D Desktop setup was significantly higher than the y- and z-errors ($p < 0.001$), prompting us to **confirm H2b**. We need to emphasize that the errors and biases are extremely minor. In fact, this median x-error$_{norm}$ for 2D Desktop (-0.01138) corresponds to just 1.13 mm, which, in terms of gross-anatomical registration accuracy, is more than sufficient.

Further, this error could possibly be ameliorated through a change in camera control for the user. It is possible that the x-error occurs due to the main camera in the 2D Desktop setup being aligned with either the x-axis (side view of the kidney) or the z-axis (front view), oriented towards the positive x-axis space. This could have caused subjects to have a bias on that axis. We explain a planned improvement of the user interface in Section 4. The x-error$_{norm}$ for 2D Desktop caused a bias (bias$_{norm}$ = 0.01146) about three times larger than the bias for VR Tabletop (bias$_{norm}$ = 0.00372) and about 7.7 times larger than VR Standup (bias$_{norm}$ = 0.00148), prompting us to **confirm H2a**.

In terms of **rotation accuracy**, subjects in both VR setups outperformed 2D Desktop subjects with median rotation accuracies of **16.3 degrees** (2D Desktop), **4.3 degrees** (VR Tabletop), and **5.0 degrees** (VR Standup) during the Ramp-Up phase. The median Plateau levels were **5.88 degrees** (2D Desktop), **3.89 degrees** (VR Tabletop), and **4.67 degrees** (VR Standup). While the slight improvement for the VR setups can likely be attributed to the learning effect (since the Plateau phase came after the Ramp-Up phase), the jump in accuracy for 2D Desktop users stands out. We assume that many subjects became more familiar with the rotation sliders over time and were able to memorize the values for each axis as all tasks in the Plateau phase were identical. **Fig 5**, J-L, shows the rather severe differences in rotation accuracy not only between the setups but also between individual axes for 2D Desktop subjects during the Plateau phase. The mean deviation around the x-axis was relatively small (**4.9 degrees**) but rather exorbitant for the y-axis at (**-54.15 degrees**). We can see a clear upward trend for y-axis as subjects progressed through the experiment and improved over time (see **Fig 5**J, middle line graph). Given these results, we **accept H1b**.

## 3.3 Completion time

Task completion time for the three setups and both phases is shown in **Fig 6** using a series of boxplots.

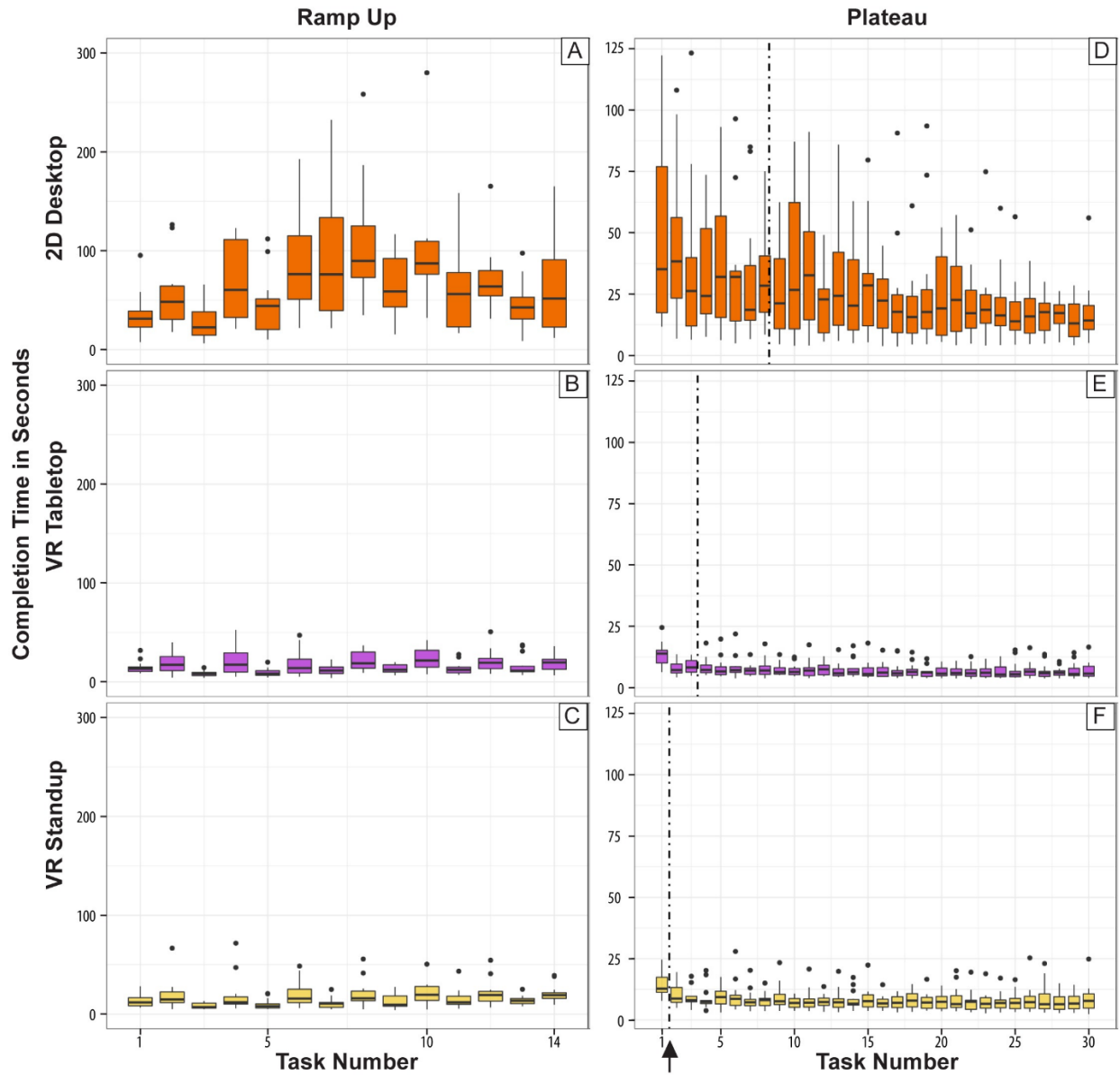

**Fig 6. Completion time for both phases and all three setups.**
(A-C) During the Ramp-Up phase. (D-F) During the Plateau phase. The vertical dash-*dot* line (black arrow) indicates after what task the plateau was reached, on average.

In the Ramp-Up phase, we found significant differences between the 2D Desktop and both VR setups but no difference between VR Tabletop and VR Standup. On average, subjects needed **67.3 seconds** for a placement task in 2D Desktop but only **16.5 seconds** in VR Tabletop and **16.3 seconds** in VR Standup, yielding a significant difference in completion time. The results of the VR setups do not differ from each other significantly. Further, in **Fig 6**A, one can clearly see the fluctuating medians for the completion time depending on the task. During odd tasks, subjects were given a prompt to focus on speed; during even tasks, we asked them to focus on accuracy. This is mirrored in the graphs for all three setups but especially so for VR Tabletop. We discuss this in more detail in Section 3.5.

Regarding RQ4, in the Plateau phase, the median Plateau level for 2D Desktop users was **22.6 seconds** after **8.3 trials** versus **7.1 seconds** after **3.43 trials** for VR Tabletop and **7.39 seconds** after just **1.5 trials** for VR Standup. Thus, it takes 2D Desktop subjects longer to reach a completion time plateau. **Fig 6**, D-F, shows the distribution of completion times during the Plateau phases. Given these findings, we **accept both H1c** (lower completion times for both VR setups) and **H4** (2D Desktop subjects need more trials to reach completion time plateau).

## 3.4 Influence of task complexity on accuracy and completion time

To answer RQ3, we computed the impact of task complexity on task accuracy and completion time during the Ramp-Up phase. **Fig 7** shows position accuracy in mm on the y-axis (i.e., centroid distance), completion time in seconds on the x-axis, and task difficulty (circle area size) for each setup.

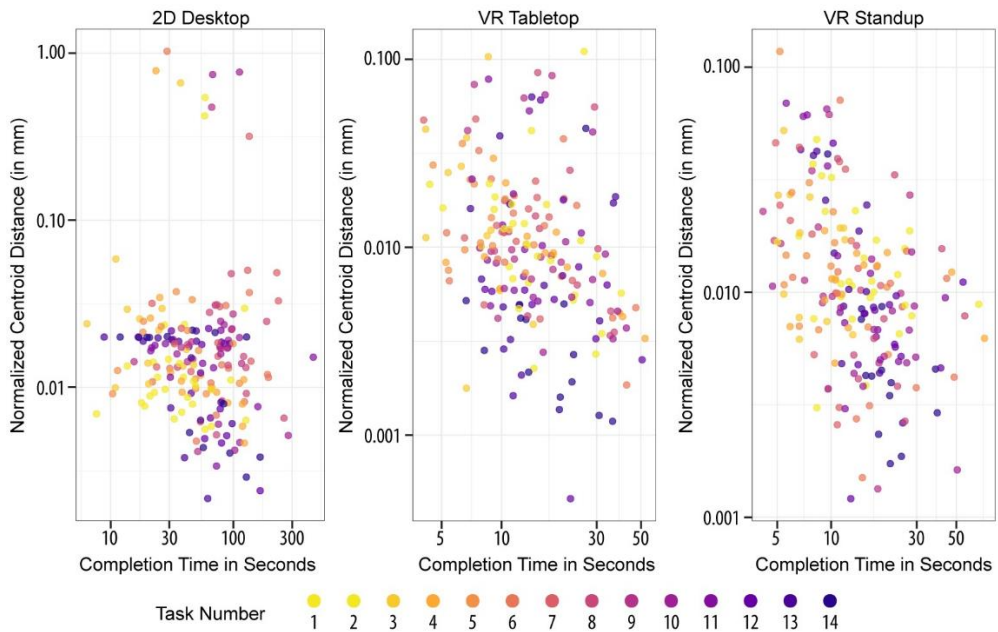

**Fig 7. Position accuracy vs. completion time dependent on task number, i.e., tissue block size, with a log-log scale.**

For VR Tabletop and VR Standup, we can see a strong cluster of records at **13.3 seconds** and **13.4 seconds**, respectively (median completion time). This is less apparent for 2D Desktop (median = 54.1 seconds) where there is more than one cluster. As becomes apparent from **Fig 7**, we found no significant correlation between task complexity and position accuracy for any setup, requiring us to **reject H3a**. We did, however, find a significant and positive Pearson correlation between task complexity and rotation accuracy, for all setups (2D Desktop: 0.457, $p < 0.001$; VR Tabletop: 0.167, $p < 0.05$; VR Standup: 0.231, $p < 0.01$). We thus **accept H3b**. Finally, for completion time, we only found a significant, positive correlation for the 2D Desktop setup (0.163, $p < 0.05$) and thus **reject H3c**.

## 3.5 Tradeoff in speed versus accuracy

Next, we wanted to understand whether there was a gain in accuracy when spending more time on a task in the Ramp-Up phase (see RQ5). Here, the results vary greatly per setup. For 2D Desktop, we found no significant Pearson correlation between completion time and any accuracy measures. For VR Tabletop, we only found a significant negative Pearson correlation between position accuracy, expressed as centroid distance (r = -0.18, p = 0.01). If controlled for instructions the subject received at the onset of the task (focus on speed vs. on accuracy), however, it becomes evident this correlation is only significant for speed tasks (r = -0.2, p = 0.05), not for accuracy tasks. Finally, for VR Standup, we identified a significant negative Pearson correlation for both position (r = -0.33, p = 0.0) and rotation accuracy (r = -0.22, p = 0.001), regardless of instructions. Given these results and the evident differences between the setups, we **reject H5**.

Note that these results are surprising as we typically see an alignment between the VR setups; however, they diverge substantially here. A possible explanation could be the larger degree of freedom for movement afforded by the VR Standup, where subjects could walk around the kidney, crouch below it if needed, and spend more time on finding a workable angle. Naturally, this setup also required the most space.

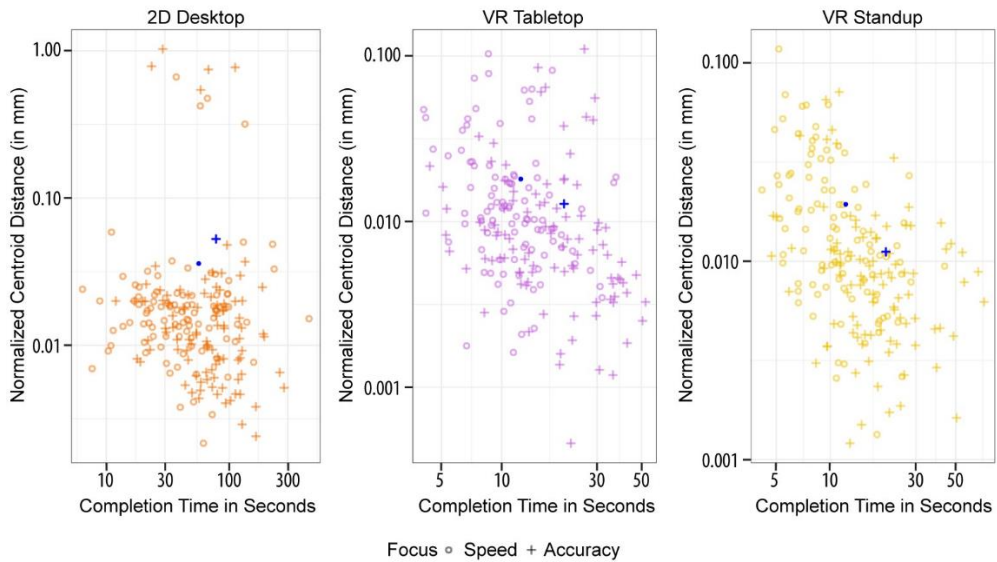

**Fig 8. Position accuracy vs. completion time dependent on instructions.** The blue circles and blue crosses mark the average completion time and position accuracy for speed and accuracy prompts, respectively.

We then performed a test to see whether subjects followed the prompts given to them when starting a new task. **Fig 8** shows position accuracy by completion time. For VR Tabletop and VR Standup, we see a tendency for longer completion times for tasks with accuracy prompts. The same pattern is evident in the boxplots in **Fig 6B** and **Fig 6C** that follow an up-and-down pattern, depending on whether the task number is odd (speed) or even (accuracy). However, none of these differences in completion time and position accuracy for the two prompts are significant, and the pattern is even less present for 2D Desktop users.

## 3.6 Satisfaction

Finally, to address RQ6, we analyzed and graphed subjects' self-reported satisfaction using the post-questionnaire data (see **Fig 9**).

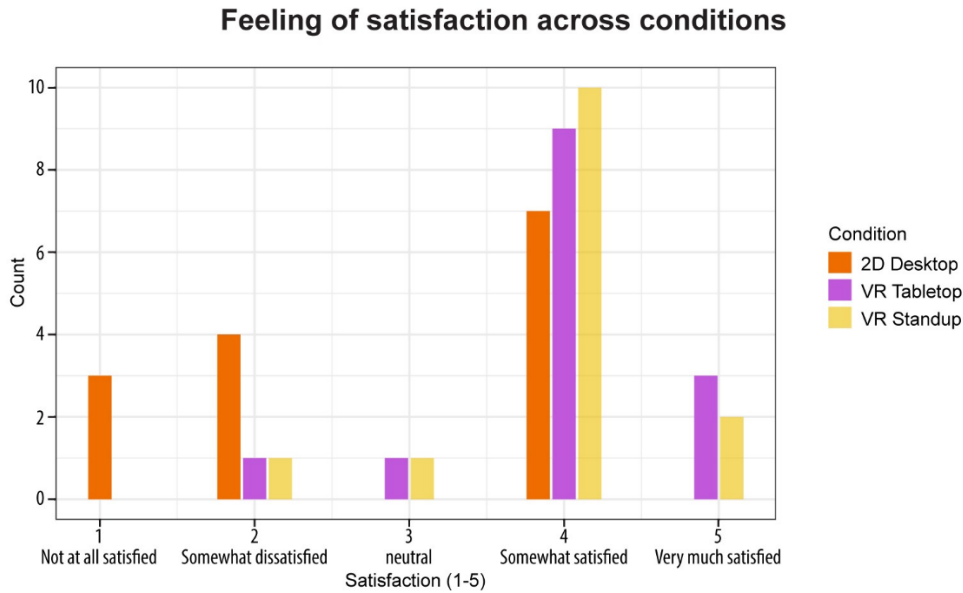

**Fig 9. Grouped bar graph of overall user satisfaction.**

Subjects used a five-point Likert scale with one (not satisfied at all) to three (neutral) to five (very much satisfied). With the overall, combined mean of **3.6** across all setups, the satisfaction was on the positive side. We then performed a pairwise Kruskal-Wallis test (with adjusted significance level for alpha inflation correction). The Kruskal-Wallis test is a non-parametric test that allows to check whether more than two non-normally distributed samples are drawn from the same distribution—i.e., it assesses whether data samples differ significantly from each other [37]. This yielded a significantly lower satisfaction for 2D Desktop users (mean = 2.79) compared to those in VR Tabletop (mean = 4) and VR Standup (mean = 3.93). The result of the VR setups does not differ significantly from each other. We thus **accept both H6a** and **H6b.** We found no correlations between satisfaction and prior experience with 3D software, first-person shooters, or VR.

## 3.7  Prior exposure to VR and 3D applications

Across the 42 subjects, there were only minor differences in previous experiences with 3D or VR applications. 24 subjects had used a VR headset before; 34 subjects had played video games in the past 12 months, 28 of whom had played first-person shooters (FPS). Further, 22 subjects had used 3D modeling software before. The largest differences appeared in previous exposure to VR between VR Standup and VR Tabletop (6 subjects vs. 10 subjects, respectively), but these did not result in a significant difference in performance. After running a comparison test, we found no differences in the distributions of completion time and accuracy measures grouping by sex, color blindness, vision impairment, age group, and right-/left-handedness. Additionally, we found no correlations between demographic variables, prior exposure to VR, or 3D applications and performance variables.

## 4. Discussion

This paper reported the results of a user study with 42 subjects involving 14 increasingly complex and 30 identical tissue block registration tasks across the 2D Desktop, VR Tabletop, and VR Standup setups. Our findings focused on comparing three different setups for the RUI in terms of **accuracy** (**position**, **rotation**), **completion time**, and **satisfaction**. Contrary to our expectations, many of our predictions were not confirmed in the study. We expected the VR Tabletop and VR Standup subjects to outperform 2D Desktop users in all of these metrics; however, we only found this to be true for rotation accuracy (**H1b**), completion time (**H1c**), and satisfaction (**H6a**), but not for position accuracy (**H1a**). From our analysis (see Sections 3.2 and 3.3), we conclude that the VR users are about **three times as fast** as Desktop users and about **a third more accurate** in terms of **rotation** for a sequence of 30 identical tasks, but **similarly accurate** for **position when normalized for the kidney height**.

We argue that several factors contributed to the high position accuracy recorded for 2D Desktop subjects:

**Restriction to two axes:** 2D Desktop users only moved the tissue block in two dimensions at a time, and the main camera was always aligned to either the x- or the z-axis. We modeled this functionality after the "quad view," which is common in 3D modeling software. It allows the user to see a 3D object from three orthographic perspectives with an additional window showing a 3D view, facilitating more precise 3D alignment. This restriction for 2D Desktop users might have played a role in their high position accuracy. In prior work, the lack of such restrictions for 3D manipulation has been shown to be a source of frustration for novice users [20]. Similarly, Masliah and Milgram [38] showed that even with advanced input devices, users separate translational (i.e., position) and rotation control when performing virtual docking tasks.

**Precision of the mouse:** The mouse proved to be a superior tool for performing fine adjustments, and the hand-eye coordination required to align the blocks seemed achievable for most subjects.

**Separate manipulation of position and rotation:** While position and rotation adjustments were performed by different tools in 2D Desktop (mouse and rotation sliders, respectively), VR Tabletop and VR Standup users performed both simultaneously (with their VR controllers). It is perfectly possible that many VR users achieved high position accuracy early on but worsened their result by subsequently adjusting the rotation of the tissue block.

Our analysis of position accuracy yielded an important insight for the continued development of the RUI. As explained in Section 3.2 and **Fig 5**, we observed an error in 2D Desktop tissue block placements, suggesting a tendency of users to place the tissue blocks according to the camera view (i.e., side or front) utilized at the time. A potential solution for this recurring error would be to implement **more than two predefined camera views**, thus giving the user multiple perspectives from which to view the reference organ.

This high position accuracy, however, was somewhat offset by the significant difference in rotation accuracy and completion time between the VR setups and the 2D Desktop setup. Yet, despite this inferiority, the 2D Desktop implementation meets the requirements outlined in Section 1.2. The tasks of the Plateau phase most closely resemble a real-world usage scenario, where multiple registrations are being performed in succession. With a median position accuracy of **1.32 mm given the kidney height on the laptop display**, 2D Desktop users got close to the goal of one mm for position accuracy. Similarly, at a median of **5.88 degrees**, the goal of rotation accuracy by 15 degrees is well met. Further, at **22.6 seconds,** the median task completion time plateau for Desktop users was within an acceptable range. In a real-world context, where the accuracy requirement is not as pronounced as it was in this study, we can expect that a reasonably accurate registration can be achieved in less time. In future studies, the research on accuracy from human tissue registration presented here will serve to support so-called Stage 2 registration at the single-cell level using image registration software and machine learning.

Finally, another goal was to make the RUI experience satisfying. In this regard, the 2D Desktop implementation was clearly lacking with a significantly lower self-reported satisfaction score than either of the VR setups (**H6a**), between which we found no significant difference (**H6b**). However, it is important to remember that this study only crudely approximates a real-world usage scenario, where the high level of accuracy and completion time suggested in this study is likely not necessary, resulting in less pressure on the user to keep adjusting their tissue blocks. This is corroborated by the fact that more time invested does not result in higher accuracy for the 2D Desktop setup (**H5**), making it more "forgiving" to users who choose to spend less time on getting a "perfect" registration. Additionally, the ease of use and wide availability of high-resolution 2D screens and computer mice is likely an advantage for users who have never experienced VR before.

Additionally, we can assume that 2D Desktop technology is less likely to cause technology frustration as 2D computer monitors, of various resolutions and size, and mice are widely available, easy to service, and use. As VR equipment becomes cheaper, less bulky, and easier to set up, VR setups may catch up, but at the time of this writing, 2D Desktop setups hold a clear advantage in this regard.

## 5. Conclusions

The insights gained in this study inform the continued development of the RUI Desktop setup as part of the HuBMAP Ingest Portal (see revised RUI 1.7 available at [https://hubmapconsortium.github.io/ccf-ui/rui/](https://hubmapconsortium.github.io/ccf-ui/rui/)). The revised RUI is optimized for Google Chrome, Firefox, and the latest (Chromium-based) version of Edge. As of August 24, 2021, there are five funded groups (including teams from outside of HuBMAP) which use the revised RUI. 82 tissue blocks were registered (26 for heart, 17 for left kidney, 15 for right kidney, 20 for spleen, and four for colon), with the potential for 147 more over the coming weeks. Further, 63 predetermined organ extraction sites were defined by seven experts for the large intestine (five), the heart (53), the left and right kidneys (one each), and the spleen (three). The 82 published tissue blocks can now be explored by anyone with an internet connection in the sister interface to the RUI, the Exploration User Interface (EUI) at [https://portal.hubmapconsortium.org/ccf-eui](https://portal.hubmapconsortium.org/ccf-eui). The average sizes for the 29 tissue blocks that were registered by HuBMAP teams for the kidney are: 21.6 mm x 13.9 mm x 4.8 mm (H x W x D) for the left kidney and 22.2 mm x 13.3 mm x 6.1 mm (H x W x D) for the right kidney. This does not contain the sizes for the kidney blocks registered by non-HuBMAP teams. Further, 63 predetermined organ extraction sites were defined by seven experts for the large intestine (five), the heart (53), the left and right kidneys (one each), and the spleen (three) and these extraction sites can be associated with hundreds of tissue blocks that share these locations.

Going forward, we envision two types of user studies exploring 3D manipulation further. First, we plan to run studies in a more "in the wild" setting [39]. This would allow us to consider variables that are hard to test in a lab setting with mostly novice users, and result in more accurate data about user performance and satisfaction in a true production setting. This would likely be a more focused study with a smaller sample of subject matter experts at their place of work (i.e., a wet lab or adjacent data processing facility), and would enable us to evaluate the performance of the 2D Desktop RUI in a realistic usage scenario.

Second, it would be valuable to test how interventions could help users improve their performance during the experiment (e.g., between the Ramp-Up and Plateau phases). Specifically, we aim to run a study with a "reflective" phase where the user sees a visualization of their own performance data from previous tasks before completing a second set of tasks. Our goal is to use the human ability to recognize patterns and trends visually to test if different types of interactive data visualizations can help users formulate strategies to improve their performance in terms of position accuracy, rotation accuracy, and completion time. Given the detailed telemetry data collected from RUI users (especially those in VR), a natural next step would be to add an intervention where users can see their own movement as well as the position and rotation of the tissue and target blocks over time, thus enabling them to detect problems and strategize more efficient solutions for future tasks. We deposited video demos and study materials for this experiment on GitHub (https://github.com/cns-iu/rui-tissue-registration).

# Acknowledgments

We would like to thank JangDong "JD" Seo from the Indiana Statistical Consulting Center for his expert input on the power analysis to determine the number of subjects needed to achieve significant results in our study; Robert L. Goldstone for providing input on the initial research design; and Leonard Cross for his expert input during the development of the three RUI setups. We further express our gratitude to Kristen Browne (Bioinformatics and Computational Biosciences Branch, Office of Cyber Infrastructure and Computational Biology, National Institute of Allergy and Infectious Diseases, National Institutes of Health) for creating the 3D model of the kidney used in this study; to Todd Theriault for copy-editing and proof-reading this paper; to Perla Brown for improving the figures; and to the eight CNS colleagues who serves as pilot testers. We would also like to acknowledge Sanjay Jain, MD, PhD, and Diane Salamon, RN, for providing the pictures of dissected kidneys featured in this paper. User studies were conducted under Indiana University IRB protocol number 1910331127.

# Supporting Information

Article title: 3D Virtual Reality vs. 2D Desktop Registration User Interface Comparison

Andreas Bueckle[1*], Kilian Buehling[2], Patrick C. Shih[3], Katy Börner[1,4]

[1] Department of Intelligent Systems Engineering, Luddy School of Informatics, Computing, and Engineering, Indiana University, Bloomington, Indiana, United States of America

[2] Research Group Knowledge and Technology Transfer, Fakultät Wirtschaftswissenschaften, Technische Universität Dresden, Germany

[3] Department of Informatics, Luddy School of Informatics, Computing, and Engineering, Indiana University, Bloomington, Indiana , United States of America

[4] Department of Information and Library Science, Luddy School of Informatics, Computing, and Engineering, Indiana University, Bloomington, Indiana, United States of America

* Corresponding author

Email: abueckle@iu.edu (AB)

The following Supporting Information is available for this article:

- Supplementary text
- Tables S1 to S2

Other supplementary materials for this manuscript include:

- Qualtrics survey PDF: https://github.com/cns-iu/rui-tissue-registration/blob/main/RUI_user_study_questionnaires.pdf
- Video demonstrations of three setups (2D Desktop, VR Tabletop, VR Standup):
    - [2D Desktop](#)
    - [VR Tabletop](#)
    - [VR Standup](#)

These materials are hosted on GitHub: https://github.com/cns-iu/rui-tissue-registration

# Power Analysis

We performed a power analysis to estimate a sample size that would give us a high chance of statistically significant results. The sample size calculation was performed using G*Power [1], an open-source statistical power analysis tool developed at the University of Duesseldorf in Germany. Among other functionality, G*Power allows the user to set a range of metrics to retrieve a sample size for a given research design as expressed through the metrics outlined below. Before being able to use G*Power, we needed to parametrize our study.

The user study presented in our paper was the first in a series of similar studies. Thus, an essential factor when parametrizing this research design for sample size calculation was that it contained two separate user studies that differed by the number of groups (user study #1 (in this paper): 3 setups = 3 groups; user study #2 (not described here): 3 setups x 2 cohorts = 6 groups), number of measurements, and thus the required sample size (see **Error! Reference source not found.**). For the subsequent user study (discussed elsewhere, in progress), we used the data from our RUI study described here as control cohort.

The statistical power calculation also needed other parameters from the researcher:

- **Test family**
    - We chose "F tests" since it allowed us to perform ANOVA for a comparison of repeated measures between groups of subjects
- **Statistical test**
    - The specific test we chose to perform was a between-factors, repeated-measures ANOVA since we aimed to analyze between subjects/users (as opposed to just within subjects—i.e., just within users)
- **Type of power analysis**

- G*Power allowed us to compute a variety of power analysis metrics. We chose "A priori: Compute required sample size" in order to determine the number of subjects needed for this study
- **Effect size**
    - The effect size is defined as the quotient of the mean difference over the standard deviation. In G*Power, the effect size can range from 0 to 50. It is always positive or zero. The default effect size in G*Power was set to 0.25. We chose 0.3 for our effect size calculation.
    - Literature [2, 3] suggests that it is common to distinguish between small (0.2), medium (0.5) and high (0.8) effect sizes in power analysis, so choosing 0.3 seemed reasonably conservative (as sample size is negatively correlated with effect size, *ceteris paribus*). Effect size is commonly seen as either small, medium, or large. Ideally, the effect size can be estimated by looking at pilot study data. The pilot study data at hand, however, may not be sufficient in order to produce an accurate result for this estimate.
- **Alpha error probability**
    - Probability of Type-I error
    - Our assumed alpha error probability of 0.05, which is a standard value.
- **Power**
    - 1 - Probability of Type-II error (beta)
    - Standard value for beta is usually set to four times the alpha error probability
    - We set the power metric to 0.8.

Given these values for the metrics required by G*Power, we obtained the results shown in **Table S1**. As mentioned above, please note that we conducted a separate user study described in a forthcoming paper where we compared two cohorts of 2D Desktop, VR Tabletop, and VR Standup users (a control and an experiment cohort). This is why we chose 6 as the "number of groups" for the Plateau phase as we were interested in comparing performances between the control and experiment cohort. We took the resulting 84 subjects and divided them by 2. The users in this study (84/2 = n = 42) then served as control cohort for the forthcoming study.

**Table S1. Metrics for sample size calculation by phase.** Notice that all the values are identical except the number of groups and measurements.

| Metrics | Ramp-Up | Plateau |
|---|---|---|
| Effect size | 0.3 | 0.3 |
| Alpha error probability | 0.05 | 0.05 |
| Power | 0.8 | 0.8 |
| Number of groups | 3 | 6 |
| Number of measurements | 14 | 30 |
| Sample size | 63 | 84 |

**Error! Reference source not found.** shows that the two studies contained in our RUI VR research design required two different sample sizes to achieve the targeted effect of size 0.3. Since we were conducting both studies together, we needed to aim for the higher number: **84 subjects, or 14 subjects per group per setup** (Control-VR Standup, Control-VR Tabletop, Control-2D Desktop, Experiment-VR Standup, Experiment-VR Tabletop, Experiment-2D Desktop).

# Data Collected from Unity

**Table S2. Telemetry data logged during experiment.**

| Name of metric | Definition | Setup |
|---|---|---|
| elapsedTime | Number of seconds since the beginning of the experiment | All |
| headsetX, headsetY, headsetZ | Position of HMD, from the origin of the virtual scene (VR only), in meters | VR Tabletop, VR Standup |
| controllerLeftX, controllerLeftY, controllerLeftZ, controllerRightX, controllerRightY, controllerRightZ, | Positions of controllers, from the origin in the virtual scene (VR only), in meters | VR Tabletop, VR Standup |
| mousepositionX, mousepositionY | The user's 2D mouse movements, in pixels | 2D Desktop |
| distance | Distance between the centroids of the two blocks, in meters (VR)/ Unity scene units (2D Desktop) | All |
| currentTissueBlock.transform.position.x, currentTissueblock.transform.position.y, currentTissueblock.transform.position.z | Current position of the tissue block | All |
| objectXLength, objectYLength, objectZLength | Current dimensions of tissue block (and target) | All |
| currentObjectRotationX, currentObjectRotationY, currentObjectRotationZ | Current x, y, z rotation of tissue block | All |

| targetRotationX, targetRotationY, targetRotationZ | Current x, y, z rotation of target block | All |
| --- | --- | --- |
| angle | Current rotational difference between tissue block and target block, in degrees (0-180) | All |
| button | Current button pressed ("grab" and "menu" in VR, "left" or "right" mouse button on 2D Desktop), or function called ("reset position", "reset rotation", etc.) | All |
| side | "Left" or "right" hand (VR), "left" or "right" mouse button (2D Desktop) | VR |
| status | "Down" (pressed), "up" (released), "dragging" (2D Desktop only) | All |

# References for Supporting Information